\documentclass[journal]{IEEEtran}
\usepackage{amsmath}
\usepackage{color,multicol}
\usepackage{cite}
\usepackage{graphicx}
\usepackage{amsmath}
\usepackage{array}
\input{psfig}
\usepackage{amssymb}

\begin{document}
\title{Adaptive Bayesian Denoising for General Gaussian Distributed (GGD) Signals in Wavelet Domain}
\author{Masoud Hashemi,~\IEEEmembership{Student Member,~IEEE}, Soosan Beheshti,~\IEEEmembership{Senior Member,~IEEE}
\thanks{Masoud Hashemi is with the Institute of Biomaterials and Biomedical Engineering (IBBME),
University of Toronto (UofT). Email: sayedmasoud.hashemiamroabadi@utoronto.ca. Soosan Beheshti is with the
Department of Electrical and Computer Engineering, Ryerson
University, Toronto, ON M5B 2K3, Canada. Email: soosan@ee.ryerson.ca.}}
\maketitle
\begin{abstract}
Optimum Bayes estimator for General Gaussian Distributed (GGD) data in wavelet is provided. The GGD distribution describes a wide class of signals including natural images. A wavelet thresholding method for image denoising is proposed. Interestingly, we show that the Bayes estimator for this class of signals is well estimated by a thresholding approach. This result analytically confirms the importance of thresholding for noisy GGD signals.
We provide the optimum soft thresholding value that mimics the behavior of the Bayes estimator and minimizes the resulting error.

The value of the threshold in BayesShrink, which is one of the most used and efficient soft thresholding methods, has been provided heuristically in the literature. Our proposed method, denoted by Rigorous BayesShrink (R-BayesShrink), explains the theory of BayesShrink threshold and proves its optimality for a subclass of GDD signals. R-BayesShrink improves and generalizes the existing BayesShrink for the class of GGD signals. While the BayesShrink threshold is independent from the wavelet coefficient distribution and is just a function of noise and noiseless signal variance, our method adapts to the distribution of wavelet coefficients of each scale. It is shown that BayesShrink is a special case of our method when shape parameter in GGD is one or signal follows Laplace distribution. Our simulation results confirm the optimality of R-BayesShrink in GGD denoising with regards to Peak Signal to Noise Ratio (PSNR) and Structural Similarity (SSIM) index.

\end{abstract}

\section{Introduction}
Data denoising via wavelet thresholding is a simple yet powerful method that has been studied and used
 in various research areas ranging from communications to biomedical signal analysis.
Wavelet shrinkage is based on
rejecting those wavelet coefficients that are smaller than a
certain value and keeping the remaining coefficients. Thus, the
problem of removing noise from a set of observed data is
transformed into finding a proper threshold for the
data coefficients. The wavelet denoising method consist of three steps: 1) applying the wavelet transform to data and calculating the wavelet coefficients; 2) applying hard or soft thresholding method; and 3) calculating the denoised signal by using the inverse wavelet transform. The pioneer shrinkage methods, such as VisuShrink and SureShrink, propose
thresholds that are functions of the noise variance and the data
length \cite{softth,visu,Donoho}. Over the past fifteen years, several
thresholding approaches such as
\cite{bayesshrink,Blu,Abramovitch,Simoncelli} have been developed.
These methods provide optimum thresholds by focusing on certain
properties of the noise-free signal, and they are proposed for
particular applications of interest.
In wavelet shrinkage methods, the noise component is usually assumed
to have a Gaussian distribution, whereas various distributions have been considered to model the noiseless coefficients.
Some of the signal classes are as follows: family of Pearson's
distributions \cite{Foucher}, Laplace distribution \cite{Sendur},
family of S$\alpha$S distributions \cite{Achim}, or Gauss-
Markov random field model \cite{Gleich}. It is important to note that, in general, the distribution of wavelet coefficients
changes from scale to scale. Two approaches have been used by different
groups to deal with this change. One approach is to use a fixed model in all scales which
decreases the modeling and denoising accuracy. The second solution
is to fit the considered  distribution
of the wavelet coefficients at each scale by optimizing the parameter in that scale \cite{Foucher,Achim,Gleich}.
However, it is known that
wavelet transform is not a complete decorrelator. Secondary properties in wavelet transform such as clustering and persistence result in Large/small values of wavelet coefficients to propagates across the different scales. To model these properties, \cite{Luettgen} and \cite{Basseville} have jointly used Gaussian model, which can easily capture the correlation between the wavelet coefficients. Nevertheless, this model is unrealistic since the wavelet coefficients have heavy tailed distribution and in addition, the proposed models are unable to include frequent sparsity and compression properties of wavelet coefficients. Consequently, wavelet based Hidden Markov models (HMM) that capture the dependencies between the coefficients with non-Gaussian distribution have been proposed \cite{Crouse}.
Although these methods outperform the thresholding methods, they are computationally expensive and time consuming which make them inefficient when speed and time are of great importance. In these cases we compromise between complexity and performance and lean towards thresholding algorithms.

In this paper we aim to find the Bayesian estimate of the noiseless signal from its
observed noisy wavelet coefficients for General Gaussian Distributed
(GGD) signals. We assume that the subbands of wavelet transformation of the
images follow General Gaussian distribution, in which the shape parameter
 is often between 0.5 and 1. This distribution is highly accepted in the image processing community for wavelet coefficients
of image subbands \cite{Ruggeri,GGD1,GGD2,GGD3,GGD4}. Our results show that the optimum Bayes estimator in this class of signals behaves similarly to a soft thresholding method. Therefore, it confirms the importance and efficiency of the existing soft thresholding approaches for image denoising purposes. One of the well known and most efficient image thresholding methods is BayesShrink \cite{bayesshrink}, which provides the threshold based on minimizing the Bayes risk $heuristically$. The threshold value proposed in BayesShrink is a function of noise variance and variance of noiseless data and is independent from the shape parameter. However, since distribution of
wavelet coefficients vary in each scale, using a threshold value which is not adapted for each scale is not very accurate. Here we propose a generalized
threshold value which adapts to distribution shape of coefficients in each scale.
Simulation results confirm that the newly proposed method denoted by Rigorous BayesShrink(R-BayesShrink) outperforms the BayesShrink itself.

The paper is organized as follows. In Section \ref{sec:Problem Formulation} the denoising problem is formulated.
In Section \ref{sec:bayesestimate} we calculate the Bayes estimator for GGD signals. The proposed thresholding algorithm based on the Bayes estimator is provided in Section \ref{lastt}. Section \ref{sec:results} contains the simulation results, and finally in Section \ref{sec:conclusion} conclusions are drawn.

\section{Problem Formulation}\label{sec:Problem Formulation}
We denote the noise-free data of length
$N$ with vector $\bar{y}^N= [\bar{y}_1,\cdots,\bar{y}_N]^T$. The noise-free data is corrupted by
an additive white Gaussian random process $w^N
=[w_1,\cdots,w_N]^T$ with zero mean and variance of $\sigma_w^2$. Therefore, the observed data $y^N=[y_1,\cdots,y_N]^T$ is
\begin{eqnarray}
y_i=\bar y_i+w_i
\end{eqnarray}
which can be expressed in terms of a desired orthonormal wavelet basis in the following form
\begin{eqnarray}
\theta_i = \bar \theta_i+v_i, \label{coef}
\end{eqnarray}
where $\theta_i$ is the $i$th noisy wavelet coefficients, $\bar \theta_i$ is the $i$th noiseless coefficient and $v_i$s are the noise wavelet coefficients. Since orthonormal bases are used, $v_i$s are also
additive white Gaussian random variables with the same
mean and variance as the noise samples ($w_i$).

We assume that the noiseless coefficients ($\bar \theta$) have zero mean General Gaussian Distribution \cite{bayesshrink},\cite{GGD1,GGD2,GGD3,GGD4}:
\begin{equation}
f(\bar{\theta})=C(\sigma_{\bar{y}},\beta)\exp\{-[\alpha(\sigma_{\bar{y}},\beta)|\bar{\theta}|]^{\beta}\}
\end{equation}
where $\beta$ is the shape parameter, $\alpha(\sigma_{\bar{y}},\beta)=\sigma_{\bar{y}}^{-1}[\frac{\Gamma(\frac{3}{\beta)}}{\Gamma(\frac{1}{\beta})}]^2$, $C(\sigma_{\bar{y}},\beta)=\frac{\beta\alpha(\sigma_{\bar{y}},\beta)}{2 \Gamma(\frac{1}{\beta})}$ and  $\Gamma(x)=\int_0^{\infty}t^{x-1}e^{-t}dt$.

When $\beta$ is one, GGD is the same as Laplacian distribution and when $\beta$ is two,
GGD describes the Gaussian distribution. The objective of this work is to provide the optimum soft threshold, $T$, that minimizes the Bayes risk. This question is addressed by BayesShrink approach heuristically and its threshold is provided by numerical analysis. Our goal is to search for the optimum threshold analytically.\\

As a reminder, in both hard (\ref{eqn:hardth}) and soft (\ref{eqn:softth}) thresholding, small absolute values of $\theta$ are mapped to zero \cite{Donoho}. The hard thresholding keeps
the values larger than the threshold. So the map is along $\hat\theta=\theta$ after that threshold.

\begin{equation}\label{eqn:hardth}
\hat{\theta}_T(i)=
\begin{cases}
  0 &  \mbox{if } \theta(i)< T, \\
  \theta(i)  & \mbox{ if } \mbox{otherwise}.
\end{cases}
\end{equation}

While the soft thresholding map is off this diagonal line by the amount of the threshold.

\begin{equation}\label{eqn:softth}
\hat{\theta}_T(i)=\text{sign}(\theta(i))(|\theta(i)|-T)_+
\end{equation}
where $(x)_+=\max(x,0)$.
%%%%%%%%%%%%%%%%%%%%%%%%%%%%%%%%%%%%%%%%%%%%%%%%%%%%%%%BayesEstimation%%%%%%%%%%%%%%%%%%%%%%%%%%%%%%%%%%%%%%

\section{Bayesian Estimate of GGD Signals}\label{sec:bayesestimate}
Here we calculate the general Bayes estimate structure for the GGD signals.
The Bayes estimate of the desired noise-free parameter $\bar \theta$, denoted by $\hat \theta$, minimizes the following mean square error (MSE)
\begin{equation}\label{eqn:bayesrisk}
\hat \theta_B= \arg_{\hat \theta} \min E(\hat{\theta}-\bar{\theta})^2
\end{equation}
This estimator is the equivalent of the mean of the posterior distribution that is an unbiased
least-squares estimate of $\bar{\theta}$, given measurement
$\theta$ \cite{Simoncelli}
\begin{equation}\label{eqn:bayes1}
\hat{\theta}_B=\int\bar{\theta}
f_{\bar{\theta}|\theta}(\bar{\theta}|\theta) \text{ d}\bar{\theta}
%\end{equation}
%\begin{equation*}
=\frac{\int\bar{\theta}
f_{\theta|\bar{\theta}}(\theta|\bar{\theta})f_{\bar{\theta}}(\bar{\theta})
\text{ d}\bar{\theta}}{\int
f_{\theta|\bar{\theta}}(\theta|\bar{\theta})f_{\bar{\theta}}(\bar{\theta})
\text{ d}\bar{\theta}}
\end{equation}
Due to the Gaussian distribution of the additive noise, we have
$$f_{\theta|\bar{\theta}}(\theta|\bar{\theta})=f_v(\theta-\bar{\theta})=\frac{1}{\sqrt{2
\pi}\sigma_w}
 e^{-\frac{(\bar{\theta}-\theta)^2}{2\sigma_w^2}}$$
and thus (\ref{eqn:bayes1}) can be written as
\begin{equation}\label{eqn:bayes2}
\hat \theta_B =\frac{\int\bar{\theta}
f_v(\theta-\bar{\theta})f_{\bar{\theta}}(\bar{\theta})
 \text{ d}\bar{\theta}}{\int f_v(\theta-\bar{\theta})f_{\bar{\theta}}(\bar{\theta}) \text{ d}\bar{\theta}}
\end{equation}
For $\beta=2$ (the Gaussian case) this Bayes estimator will become
\begin{equation}
\hat{\theta}_B=\frac{\sigma^2_{\bar{y}}}{\sigma_{\bar{y}}^2+\sigma_w^2}\theta
\end{equation}
where $\sigma_{\bar y}^2$ is the variances of noiseless data.

 While equation (\ref{eqn:bayes2}) cannot be solved for all other
values of $\beta$, it will have the following analytical form for $\beta=1$ (the Laplacian case)
\begin{multline}
\hat{\theta}_B=
\frac{e^{\frac{\sqrt{2}\theta}{\sigma_{\bar
y}}}(2\sigma_w^2+\sqrt{2}\sigma_{\bar y} \theta)
Q(\frac{\sqrt{2}\sigma_w}{\sigma_{\bar y}}+\frac{\theta}{\sigma_w})}
{\sqrt{2}\sigma_{\bar y}[e^{\frac{\sqrt{2}\theta}{\sigma_{\bar y}}}Q(\frac{\sqrt{2}\sigma_w}{\sigma_{\bar y}}+
\frac{\theta}{\sigma_w})+e^{\frac{-\sqrt{2}\theta}{\sigma_{\bar
y}}}Q(\frac{\sqrt{2}\sigma_w}{\sigma_{\bar y}}
-\frac{\theta}{\sigma_w})]}
\\
-
\frac{e^{-\frac{\sqrt{2}\theta}{\sigma_{\bar
y}}} (2\sigma_w^2-\sqrt{2}\sigma_{\bar y} \theta)
Q(\frac{\sqrt{2}\sigma_w} {\sigma_{\bar y}}-\frac{\theta}{\sigma_w})}
{\sqrt{2}\sigma_{\bar y}[e^{\frac{\sqrt{2}\theta}{\sigma_{\bar y}}}Q(\frac{\sqrt{2}\sigma_w}{\sigma_{\bar y}}+
\frac{\theta}{\sigma_w})+e^{\frac{-\sqrt{2}\theta}{\sigma_{\bar
y}}}Q(\frac{\sqrt{2}\sigma_w}{\sigma_{\bar y}}
-\frac{\theta}{\sigma_w})]} \label{11}
\end{multline}

in which $Q(x)=\frac{1}{\sqrt{2\pi}}\int_x^{\infty}e^{\frac{-t^2}{2}}dt$. Figure \ref{fig:DiffSigmay}, shows this Bayes estimator, $\hat{\theta}_B$, for
different values of $\sigma_{\bar{y}}$ while $\sigma_w$ is constant and

\section{Mean Square Error(MSE) Soft Threshold Fitting for the Bayes Estimator} \label{lastt}
\begin{figure}[t]
  \centering
  \includegraphics[width=9cm]{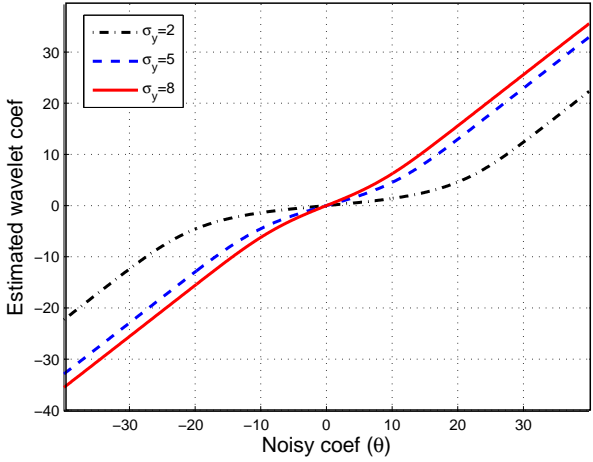}\\
  \caption{Bayes estimator for different noiseless data variance $\sigma_{\bar{y}}$ while noise variance is $\sigma_{w}$ is five and shape parameter $\beta$ is one.}
\label{fig:DiffSigmay}
\end{figure}

\begin{figure}[t]
  \centering
  \includegraphics[width=9cm]{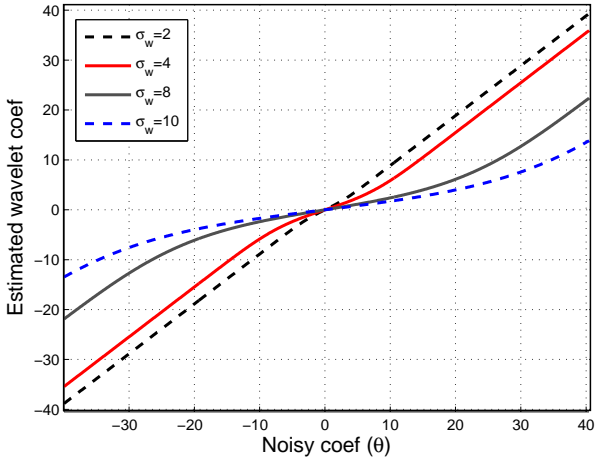}\\
  \caption{Bayes estimator curves for different noise variance $\sigma_{w}$ while noiseless data variance $\sigma_{\bar{y}}$ is five and   shape parameter $\beta$ is one.}
\label{fig:DiffSigman}
\end{figure}

We studied the behavior of the Bayes estimator of
noisy GGD signals, $\hat\theta_B$ in (\ref{eqn:bayes2}), as the noise variance, the noise-free variance and the shape parameter vary. Our analysis confirms that the estimator behaves similarly to a soft thresholding algorithm \ref{eqn:softth} as will be described.  Figures \ref{fig:DiffSigmay} and \ref{fig:DiffSigman} show examples of variation of this estimator.

\begin{figure}[t]
  \centering
  \includegraphics[width=9cm]{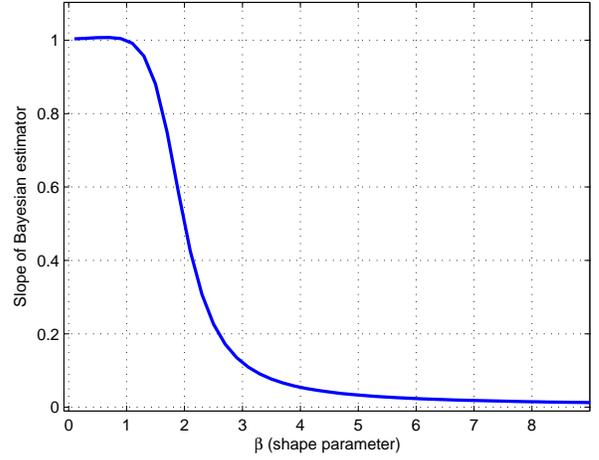}\\
  \caption{Asymptotic slope of the Bayesian estimator for large coefficients as a function of the shape parameter $\beta$.} \label{fig:slopes}
\end{figure}

\begin{figure*}[t]
  \centering
  \includegraphics[width=18cm]{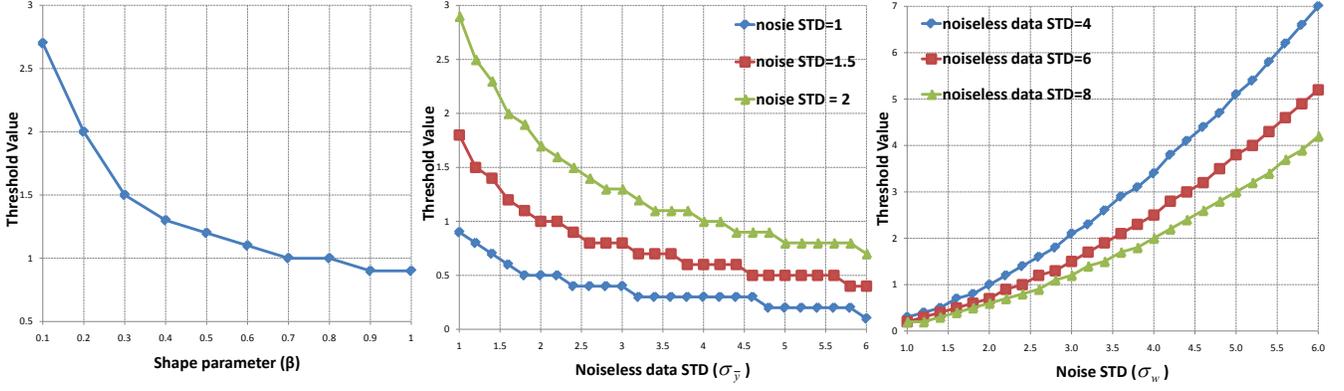}\\
  \caption{Left: Threshold values ($T^*$) versus $\beta$ when $\sigma_{\bar{y}}$ and $\sigma_{w}$ are constant;
           Middle: Threshold values ($T^*$) versus $\sigma_{\bar{y}}$ when $\sigma_{w}$ and $\beta$ are constant;
           Right: Threshold values ($T^*$) versus $\sigma_{w}$ when $\sigma_{\bar{y}}$ and $\beta$ are constant.}
\label{fig:changeparameterbayesian}
\end{figure*}

Note that a soft thresholding map is an odd function of the noisy coefficient and is zero up to a particular point $T$ and then follows a line with slope of one. Our observation, as depicted in these two figures, confirms that the
Bayes estimate is an odd function of the noisy coefficients that is zero at zero, grows nonlinearly with a small amount to a particular point and after that it asymptotically follows a line for larger wavelet coefficients. In Figure \ref{fig:slopes} we show the values of this slope as the shape parameter $\beta$ varies. As the figure shows, the slope is extremely close to one for when the same parameter is between zero and one. This behavior mimics the soft thresholding. It is known that in natural images we deal only  with the shape parameters between 0.5 to 1 \cite{bayesshrink,GGD1,GGD2,GGD3,GGD4}. Therefore, a soft thresholding estimate of the Bayes estimate can very well mimic the behavior of this estimator \footnote{Note that for the $\beta$'s out of this range, we still can use weighted version of the coefficients based on Figure \ref{fig:slopes} and then provide a thresholding.}.
To find the optimum thresholding mapping, we minimize the following least square error between the Bayes estimate in (\ref{eqn:bayes2}) and the soft thresholding by value $T$ in (\ref{eqn:softth}):
\begin{equation} \label{eqn:Tpara}
T^*= \arg \min_T \int_{-\infty}^{+\infty} f(\bar{\theta}) (\hat{\theta}_B-\hat{\theta}_T)^2 \text{d}\theta
\end{equation}
Figure \ref{fig:changeparameterbayesian} shows these threshold values found for a range of noise variance, signal variance, and shape parameter. Our study, as depicted in this figure, resulted in the following choice of a closed form representation that best describes this threshold value as a function of noise variance ($\sigma_w$), noiseless variance ($\sigma_{\bar{y}}$), and shape parameter ($\beta$):
 \begin{equation}
 T=a  \beta^{b_1} \sigma_w^{b_2(\beta)}\sigma_{\bar{y}}^{b_3(\beta)}, \label{varr}
 \end{equation}
 where parameters $a$, $b_1$, $b_2(\beta)$ and $b_3(beta)$ can be calculated through another least square optimization. The following optimum values have been found by solving this optimization problem over a feasible range of noise variance, signal variance, and shape parameter:
 \begin{equation}
a=1, \; b_1=-\frac{1}{2},\; b_2=1+\sqrt{\beta}, \; b_3=-\sqrt{\beta}
 \end{equation}
The above parameters lead to the following optimum soft threshold that best mimics the Bayes estimator
\begin{equation}\label{eqn:proposed}
T_{R} = \frac{1}{\sqrt{\beta}} \sigma_w (\frac{\sigma_w}{\sigma_{\bar{y}}})^{\sqrt{\beta}}
\end{equation}
We denote this threshold as Rigorous BayesShrink (R-BayesShrink).
\subsection{BayesShrink and Rigorous BayesShrink}
We showed in the previous sections that the Bayes estimator for GGD signals behaves very much like a soft thresholding algorithm. Therefore, we have shown rigorously that soft thresholding can indeed capture the behavior of Bayes estimator in this scenario. BayesShrink is one of the most efficient existing methods of wavelet thresholding for the purpose of image denoising \cite{bayesshrink}. The method searches for the optimum soft threshold that minimizes the MSE and maps the observation to a class of thresholds in the form of $\sigma_{w}^{c_1}\sigma_{\bar{y}}^{c2}$. Its numerical analysis finds the following threshold
\begin{eqnarray}
T_B = \frac{\sigma_w^2}{\sigma_{\bar{y}}} \label{BB}
 \end{eqnarray}
  Nevertheless, if we set the value of $\beta$ to one in (\ref{eqn:proposed}), we get the same threshold of the BayesShrink. This confirms and rigorously proves the importance of BayesShrink for the Laplacian case. However, as the BayesShrink threshold ignores the role of $\beta$ of the considered GGD model, it is expected that R-BaysShrink outperform the BayesShrink. Examples in our simulation results confirm that R-BayShrink indeed outperforms BayesShrink.
\subsection{Least Square Estimator(LSE) vs. Rigorous BayesShrink}
The optimization for soft threshold mapping in R-BayesShrink is based on minimizing the MSE in (\ref{eqn:Tpara}). In an earlier work, we have optimized this mapping by LSE minimization \cite{sips}
\begin{equation} \label{eqn:Tpara2}
T_{LSEB}= \arg \min_T \int_{-\infty}^{+\infty} |\hat{\theta}_B-\hat{\theta_T}|^2 \text{d}\theta
\end{equation}
 It was shown that using the structure in (\ref{varr}), the optimum threshold is
 \begin{eqnarray}
 T_{LSEB}= \frac{\sqrt{2} \beta^{1.8}\sigma_w^2}{\sigma_{\bar{y}}^{\beta}} \label{LSEB}
 \end{eqnarray}
We denote this threshold as LSE Based (LSEB).

On the other hand, the Maximum a posterior (MAP) estimator for the Laplace distribution ($\beta=1$) is
\begin{equation} \label{eqn:MAP}
\hat{\theta}_{MAP}= \arg \max_{\bar{\theta}}[\log(f_v(\theta-\bar{\theta}))+\log(f_{\bar{\theta}}(\bar{\theta}))]
\end{equation}
and since the additive noise follows a Gaussian distribution, the MAP estimate is
\begin{equation}\label{eqn:MAP2}
\hat{\theta}_{MAP}=\arg \max_{\bar{\theta}}[-\frac{(\theta-\bar{\theta})^2}{2\sigma_w^2}-log(\sigma_w\sqrt{2})-\frac{\sqrt{2}|\bar{\theta}|}{\sigma_{\bar{y}}}]
\end{equation}
%To solve (\ref{eqn:MAP2}), we should solve %$\frac{y-\hat{\theta}}{\sigma_w^2}-\frac{\sqrt{2}}{\sigma_{\bar{y}}}\text{sign}(\theta)=0$.
Solving this optimization, leads to the following soft threshold value \cite{Sendur}
\begin{eqnarray}
T_{MAP}=\frac{\sqrt{2}\sigma_w^2}{\sigma_{\bar{y}}}
\end{eqnarray}
This value is the same as the LSE based threshold mapping in (\ref{LSEB}) when the $\beta$ is set to one. This confirms that
for the Laplacian distibution, the LSE based estimator is the same as the MAP estimator! Note that this threshold value is the same as BayesShrink (R-BayesShrink for $\beta=1$) with an extra term $\sqrt{2}$.
%---------------------------------------------------------------------------

%---------------------------------------------
\subsection{Parameter Estimation of GGD}
The proposed R-BayesShrink threshold (\ref{eqn:proposed}) has three parameters: \\
1) \emph{noise variance} ($\sigma_w$): Median Absolute Deviation (MAD) estimate is the standard noise variance estimator \cite{softth,mallat}:
\begin{equation}\label{eqn:mad}
   \hat{\sigma}_{MAD}=\frac{Median(|\theta|)}{0.6745},\theta \in \; subband \;\;HH_1
\end{equation}
where $HH_1$ is the high frequency subband.

2) \emph{Noiseless data variance} ($\sigma_{\bar{y}}$):
Since the data is independent from noise, the variance of the noisy data is $$\sigma_y^2=\sigma_{\bar{y}}^2+\sigma_w^2$$
On the other hand, variance of
the noisy data can be estimated by:
\begin{equation}\label{eqn:varnoisy}
\hat{\sigma}_y^2=\frac{1}{M}\sum_{y_i \in N(k)} (y_i-m_y)^2
\end{equation}
where $M$ is the number of coefficients in scale $N(k)$ and $m_y$ is the mean in the same scale $m_y=\frac{\sum{y_i}}{M}$. Using the noise variance estimate in (\ref{eqn:mad}) and noisy data variance estimate in (\ref{eqn:varnoisy}), the estimate of
noiseless data variance is
\begin{equation}\label{eqn:varnoiseless}
\hat{\sigma}_{\bar{y}}=\sqrt{\hat{\sigma}_y^2-\hat{\sigma}_w^2}
\end{equation}
3) \emph{Shape parameter} ($\beta$):
There are three methods available for $\beta$ estimation in
Generalized Gaussian Distributed signals: A) Moment Matching
Estimator (MME), which uses moments of GGD signals; B) Entropy
matching Estimator (EME) that relies on matching the entropy of
the GGD modeled distribution with that of a set of empirical data;
and C) Maximum Likelihood Estimator (MLE), which finds the
shape parameter which maximizes the likelihood \cite{shapeparameter}.\\
MME is an accurate and still a simple method that is useful when
$\beta$ is between 0.18 and 2 which covers the range for images\cite{shapeparameter}.
Therefore, we use MME for GGD parameter estimation.

To estimate $\beta$, MME uses the second and fourth moments of the subbands.
For the Kurtosis of GGD, $\kappa_{y|y_i \in N(k)}$, we have \cite{bayesshrink}
\begin{equation}
 \kappa_{y|y_i \in N(k)}=\frac{1}{\sigma_{y}^4}(6\sigma_w^2\sigma_{y}^2-3\sigma_w^4+(\sigma_{y}^2-\sigma_w^2)^2\frac{\Gamma(\frac{1}{\beta})\Gamma(\frac{5}{\beta})}
             {\Gamma^2(\frac{3}{\beta})}) \label{ka}
\end{equation}
and $\kappa_{y|y_i \in N(k)}$ itself
can be estimated from the observed noisy data $\{y_i|y_i \in N(k),i=1,...,M\}$ as follows \cite{kurtosis}:
\begin{equation}
 \hat{\kappa}_{y|y_i \in N(k)}=\frac{\sum_{y_i \in N(k)}(y_i-m_y)^4}{(M-1)\sigma_{y}^4} \label{k}
\end{equation}
Where $m_y$ is the mean value of available $y_i$s ($m_y=\frac{\sum{y_i}}{M}$, $M$ is the number of coefficients in $k^{th}$ scale and $\hat{\sigma}_y$ is from (\ref{eqn:varnoisy})).

 Having the estimate of $\kappa_{y|y_i \in N(k)}$ from (\ref{k}), estimate of noise variance from (\ref{eqn:mad}), and
 $\sigma_{\bar{y}}$ estimate from (\ref{eqn:varnoiseless}), we can estimate $\beta$ from (\ref{ka}). Table \ref{tab:betaest} shows some of the estimated $\beta$'s from the given noisy data. It is evident that the estimated values are reasonably accurate.

%---------------------------------------PSNR when 500mA is used as ref-------------------------
\begin{table}
 \begin{center}
  \caption{Shape parameter ($\beta$) estimated from different noisy data with different SNR's using MME.}\label{tab:betaest}
  \begin{tabular}{ | c | c | c | c | c | } \hline
  $\beta$ & 0.4  & 0.6  & 0.8  & 1.0  \\ \hline
   SNR=15 & 0.45 & 0.69 & 0.83 & 1.01 \\ \hline
   SNR=30 & 0.41 & 0.68	& 0.83 & 1.03 \\ \hline
  \end{tabular}
 \end{center}
\end{table}
%--------------------------------------------------------------------------

%--------------------------------------------------------------------------
\section{Simulation Results} \label{sec:results}
Here we compare the performance of our proposed algorithm and BayesShrink for three sets of data:\\
1) Synthetic data, which is a set of GGD data with a known shape parameter.\\
2) Natural images.\\
3) Computed Tomographic (CT) images.\\
In all the tests, we use Haar wavelets in 5 levels. Each test is run one hundred times and image sizes are $512 \times 512$.

To measure the quality of the denoised images, Peak Signal to Noise Ratio (PSNR) is used which is defined as follows:
\begin{equation} \label{eqn:psnr}
{\rm PSNR}=10 \log_{10} \frac{(2^B-1)^2}{\rm MSE}
\end{equation}
 where $B$ is the number of bits in the images and MSE is the Mean Square Error (MSE). For natural images $B=8$ and for CT phantom images $B=16$.\\
Table \ref{tab:synthtest} shows the PSNR of denoised synthetic data with BayesShrink in (\ref{BB}), LSEB in (\ref{LSEB}), and R-BayesShrink in (\ref{eqn:proposed}). Based on this table, R-BayesShrink outperforms the other two methods and as was expected, for $\beta=1$, R-BayesShrink and BayesShrink performance are the same.\\
%---------------------------------------------------------------------------------
\begin{table}
 \begin{center}
  {\footnotesize
  \caption{PSNR of synthetic data denoised with BayesShrink, LSEB, and R-BayesShrink.}\label{tab:synthtest}
  \begin{tabular}{ | c | c | c | c | c | c | c | } \hline

SNR	& 5	& 10 & 15& 20& 25& 30\\ \hline
        \multicolumn{7}{|c|}{$\beta$=0.1} \\ \hline 						
Bayesshrink	&  35.22& 37.29& 45.36& 49.77& 52.22& 59.19\\ \hline
LSEB&           36.63& 38.24& 45.82& 50.02& 52.35& 59.27\\ \hline
R-BayesShrink& 37.77& 44.97& 49.23& 53.10& 55.17& 63.10\\ \hline
        \multicolumn{7}{|c|}{$\beta$=0.3} \\ \hline 				
Bayesshrink&   33.15& 37.56& 41.68& 44.63& 52.46& 56.85\\ \hline
LSEB&           33.83& 37.99& 41.91& 44.75& 52.51& 56.87\\ \hline
R-BayesShrink& 34.08& 38.47& 42.41& 45.18& 52.77& 57.04\\ \hline
        \multicolumn{7}{|c|}{$\beta$=0.5} \\ \hline 					
Bayesshrink  & 27.74& 30.73& 38.45& 40.43& 43.92& 47.40\\ \hline
LSEB         & 27.97& 30.92& 38.56& 40.47& 43.95& 47.40\\ \hline
R-BayesShrink& 27.97& 30.98& 38.65& 40.50& 43.97& 47.41\\ \hline
        \multicolumn{7}{|c|}{$\beta$=0.8} \\ \hline 				
Bayesshrink&   21.03& 24.52& 33.28& 33.84& 39.28& 45.07\\ \hline
LSEB         & 20.98& 24.55& 33.31& 33.85& 39.29& 45.08\\ \hline
R-BayesShrink& 21.05& 24.56& 33.31& 33.85& 39.29& 45.10\\ \hline
        \multicolumn{7}{|c|}{$\beta$=1} \\ \hline 						
Bayesshrink&   18.99& 23.26& 29.36& 33.71& 40.18& 44.58\\ \hline
LSEB&           18.80& 23.24& 29.35& 33.70& 40.17& 44.56\\ \hline
R-BayesShrink& 18.99& 23.26& 29.36& 33.71& 40.18& 44.58\\ \hline

 \end{tabular}}
 \end{center}
\end{table}
%----------------------------------------------------------------------------------
Since R-BayesShrink performs better than LSEB, we eliminate the comparison of LSEB in the following examples.

For the natural images (CameraMan, Lena, Mandrill, peppers, goldhill, and boatshown) the denoising results are given
in Table \ref{tab:ch5result}.

\begin{figure*}
  \centering
  \includegraphics[width=12cm]{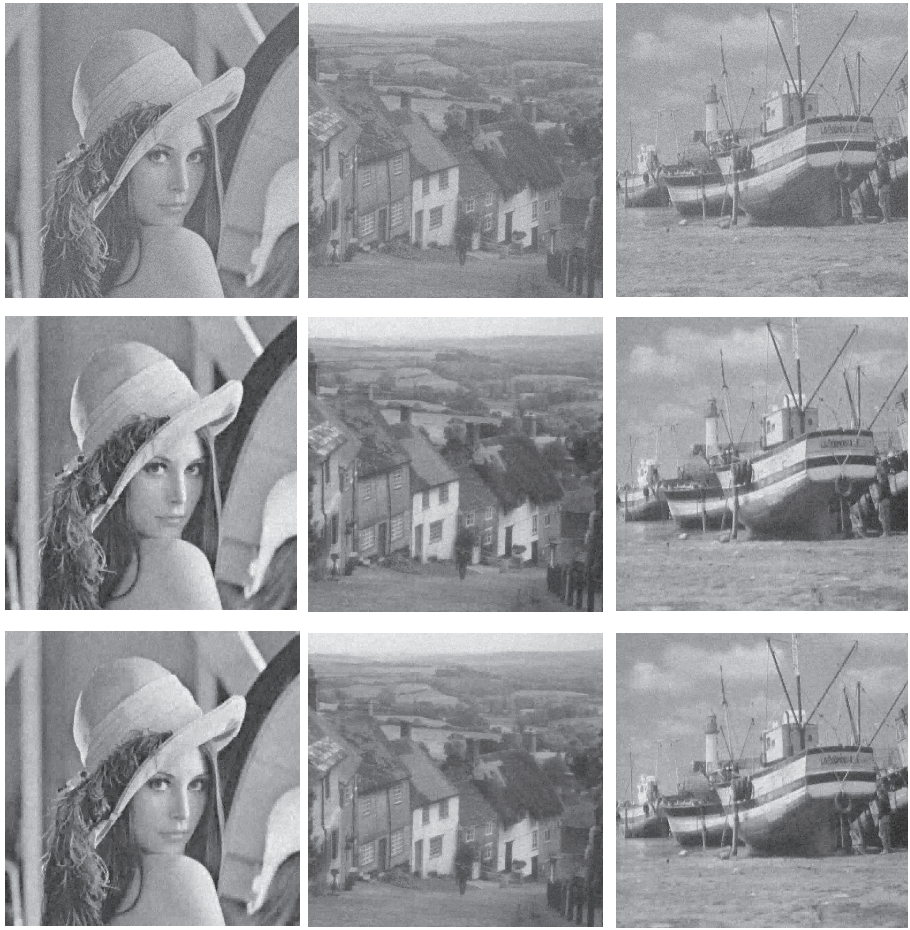}\\
  \caption{Denoised images with BayesShrink and R-BayesShrink. Top: Noisy images with SNR=15, middle: denoised with BayesShrink and bottom: denoised by R-BayesShrink.}
\label{fig:SNR15Res}
\end{figure*}

%-----------------------------PSNR of natural images---------------------------
\begin{table}
 \begin{center}
  {\footnotesize
  \caption{PSNR of denoised images with BayesShrink and the proposed threshold (R-BayesShrink).}\label{tab:ch5result}
  \begin{tabular}{ | c | c | c | c | c | c | c | } \hline

       SNR        & 5     & 10    & 15    & 20    & 25    & 30    \\ \hline
                \multicolumn{7}{|c|}{CameraMan} \\ \hline
      BayesShrink & 20.99 & 23.14 & 25.92 & 29.25 & 32.99 & 37.18 \\ \hline
    R-BayesShrink & 21.08 & 23.24 & 26.09 & 29.41 & 33.20 & 37.28 \\ \hline

         \multicolumn{7}{|c|}{Lena} \\ \hline
      BayesShrink & 23.89 & 25.89 & 28.14 & 30.78 & 33.82 & 37.35 \\ \hline
    R-BayesShrink & 23.94 &	25.96 & 28.27 &	30.91 &	34.00 &	37.54 \\ \hline

         \multicolumn{7}{|c|}{Mandrill} \\ \hline
      BayesShrink & 20.04 &	21.52 &	23.94 &	27.40 &	31.46 & 35.98 \\ \hline
    R-BayesShrink & 20.07 &	21.57 &	23.94 & 27.41 & 31.48 &	36.00 \\ \hline

         \multicolumn{7}{|c|}{GoldHill} \\ \hline
      BayesShrink & 24.00 &	25.68 &	27.67 &	30.38 &	33.37 & 37.42 \\ \hline
    R-BayesShrink & 24.04 &	25.71 &	27.73 &	30.38 & 33.54 &	37.48 \\ \hline

         \multicolumn{7}{|c|}{Peppers} \\ \hline
      BayesShrink & 20.78 &	23.29 &	26.34 &	29.44 &	32.94 &	36.87 \\ \hline
    R-BayesShrink & 20.83 &	23.37 &	26.41 &	29.76 &	33.16 &	37.00 \\ \hline

         \multicolumn{7}{|c|}{Boat} \\ \hline
      BayesShrink &  22.48 & 24.35 & 26.68 & 29.42 & 32.64 & 36.45 \\ \hline
    R-BayesShrink &  22.52 & 24.43 & 26.77 & 29.55 & 32.80 & 36.57 \\ \hline

 \end{tabular}}
 \end{center}
\end{table}
%--------------------------------------------------------------------------
 Table \ref{tab:ch5result} shows that R-BayesShrink has better PSNR values which shows its better performance in comparison with BayesShrink. Figure \ref{fig:SNR15Res} shows the denoised images with BayesShrink and R-BayesShrink. It is evident that R-BayesShrink preserves the edges better than BayesShrink. To quantitatively show this property, Structural Similarity (SSIM) index \cite{SSIM} is used, and the results are given in Tabel \ref{tab:SSIM}. SSIM index compares local patterns of pixel intensities which are  normalized for luminance and contrast.

%-----------------------------SSIM of natural images---------------------------
\begin{table}
 \begin{center}
  {\footnotesize
  \caption{SSIM of denoised images with BayesShrink and the proposed threshold (R-BayesShrink).}\label{tab:SSIM}
  \begin{tabular}{ | c | c | c | c | c | c | c | } \hline

       SNR        & 5     & 10    & 15    & 20    & 25    & 30    \\ \hline
                \multicolumn{7}{|c|}{CameraMan} \\ \hline
      BayesShrink & 0.13 & 0.30 & 0.54 & 0.77 & 0.90 & 0.97 \\ \hline
    R-BayesShrink & 0.14 & 0.33 & 0.59 & 0.83 & 0.96 & 0.99 \\ \hline

         \multicolumn{7}{|c|}{Lena} \\ \hline
      BayesShrink & 0.33 & 0.60 & 0.81 & 0.94 & 0.98 & 0.99 \\ \hline
    R-BayesShrink & 0.35 & 0.60 & 0.83 & 0.95 &	0.99 & 0.99 \\ \hline

         \multicolumn{7}{|c|}{Mandrill} \\ \hline
      BayesShrink & 0.35 &	0.74 & 0.93 & 0.993 & 0.998 & 0.999 \\ \hline
    R-BayesShrink & 0.45 &	0.79 & 0.95 & 0.996 & 0.999 & 0.999 \\ \hline

         \multicolumn{7}{|c|}{GoldHill} \\ \hline
      BayesShrink & 0.32 &	0.63 &	0.85 &	0.95 & 0.993 & 0.999 \\ \hline
    R-BayesShrink & 0.35 &	0.65 &	0.89 &	0.97 & 0.996 & 0.999 \\ \hline

         \multicolumn{7}{|c|}{Peppers} \\ \hline
      BayesShrink & 0.14 &	0.33 &	0.58 &	0.79 &	0.92 &	0.97 \\ \hline
    R-BayesShrink & 0.16 &	0.35 &	0.63 &	0.85 &	0.96 &	0.99 \\ \hline

         \multicolumn{7}{|c|}{Boat} \\ \hline
      BayesShrink &  0.34 & 0.64 & 0.85 & 0.95 & 0.990 & 0.998 \\ \hline
    R-BayesShrink &  0.37 & 0.65 & 0.87 & 0.98 & 0.995 & 0.999 \\ \hline

 \end{tabular}}
 \end{center}
\end{table}
%--------------------------------------------------------------------------

The third test is done with CT images with different x-ray doses. In these images the x-ray tube voltage is kept at 120kV, and its current (mA) is changed.
The quality of the CT images is a function of x-ray tube current and voltage. Decreasing the tube current causes a decrease in the number of photons hitting the
detectors and thereby decreases the quality and SNR. Here we test the images with 8 different currents from 10mA to 500mA. To calculate the PSNR we use the images captured by 500mA as the reference and compare the other images with it.
Table \ref{tab:CTnoise}, shows the amount of noise in the original image, the image denoised by BayesShrink, and the image denoised by R-BayesShrink. To measure the noise, we use the
variance of a smooth part of the phantom. These numbers are the evaluation measure for the amount of noise used by radiologists.

\begin{figure*}
  \centering
  \includegraphics[width=12cm]{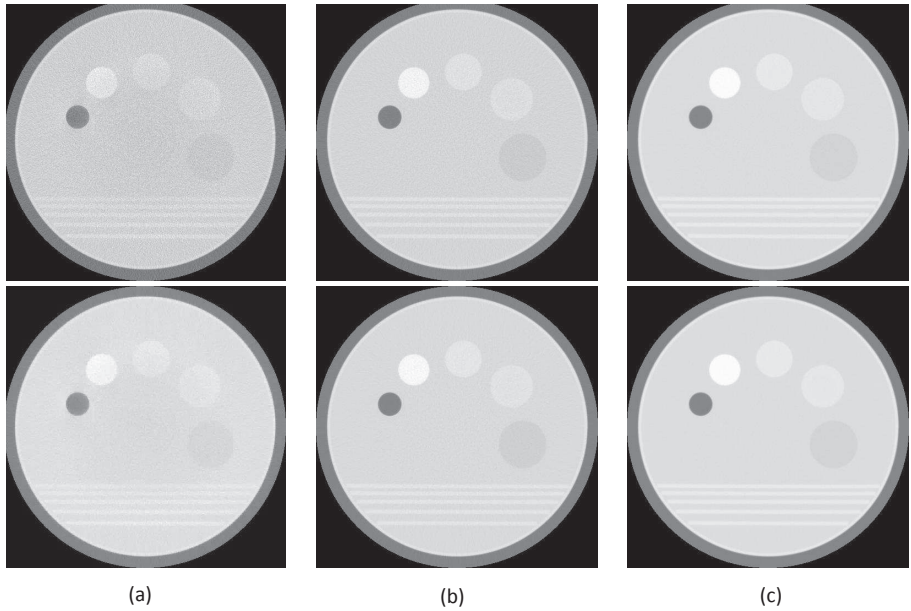}\\
  \caption{Phantom images denoised with R-BayesShrink. From left to right: 10mA, 50mA, and 300mA (Top: noisy image; Bottom: denoised image by R-BayesShrink.}
\label{fig:CTres}
\end{figure*}

%----------------------------------Noise Comparison-------------------------
\begin{table}
 \begin{center}
  {\footnotesize
  \caption{Noise standard deviation after and before denoising with BayesShrink and R-BayesShrink which is calculated based on signal variation in a smooth region (this is how radiologists calculate the amount of noise).}\label{tab:CTnoise}
  \begin{tabular}{ | @{}l | c@{} |c@{} |c@{} |c @{}|c @{}| c @{}|c @{}|c @{}|} \hline
Tube Cur. (mA)& 10 & 20 & 50 & 100 & 200 & 300 & 400 & 500 \\ \hline
Original & 79.70 & 59.50 & 29.10 & 18.80 & 12.50 & 10.20 & 9.10 & 7.62 \\ \hline
Bayes Shrink & 37.31 & 35.93 & 22.40 & 15.78 & 11.30 &  9.22 & 8.33 & 7.10 \\ \hline
R-Bayes Shrink & 19.64 & 19.74 & 14.55 &  9.35 &  7.80 &  6.55 & 5.75 & 5.10 \\ \hline
  \end{tabular}}
 \end{center}
\end{table}
%--------------------------------------------------------------------------
This table shows that R-BayesShrink reduces the noise more effectively. Figure \ref{fig:CTres} shows the CT phantom images denoised with R-BayesShrink to compare the
edges with the original one. All the edges, even the low contrast ones, are kept very well.
Table \ref{tab:CTres} compares the PSNR's of R-BayesShrink and BayesShrink. To calculate PSNR, 500mA image is used as the reference.
This table confirms the strength of R-BayesShrink in comparison with
BayesShrink.

%---------------------------------------PSNR when 500mA is used as ref-------------------------
\begin{table}
 \begin{center}
  {\footnotesize
  \caption{PSNR of denoised CT phantom with BayesShrink and R-BayesShrink.}\label{tab:CTres}
  \begin{tabular}{ | @{}l | c @{}| c@{} | c@{} | c @{}| c @{}| c@{} | c@{} | } \hline
   Tube Current(mA)& 10 & 20 & 50 & 100 & 200 & 300 & 400 \\ \hline

   BayesShrink & 14.63 & 16.32 & 20.18 & 22.84 & 24.36 & 26.52 & 26.84 \\ \hline
 R-BayesShrink & 16.14 & 18.98 & 22.31 & 24.91 & 25.46 & 27.60 & 27.72 \\ \hline

  \end{tabular}}
 \end{center}
\end{table}
%--------------------------------------------------------------------------

\section{Conclusion} \label{sec:conclusion}
We presented a wavelet based denoising method based on the structure of the Bayes estimator for Generalized Gaussian Distributed (GGD) data. We showed that for this class of signals, the Bayes estimate can be well approximated by a soft thresholding map. The optimum wavelet coefficient soft thresholding map denoted by R-BayesShrink was provided. Our proposed threshold value is a generalized form of the BayesShrink method. While the threshold value of the BayesShrink is only a function of noise variance and noise-free data variance, our proposed threshold generalizes BayesShrink by considering the shape parameter of the subbands as well. Consequently it was shown that BayesShrink is a special case of R-BayesShrink when data has a Laplacian distribution. Finally we evaluated the proposed method with three different data-sets and in all cases the proposed method showed superiority by simultaneously providing the better PSNR and better edge preservation.

\end{document}